Comb-mode-resolved adaptive sampling terahertz dual-comb spectroscopy with a free-running single-cavity fiber laser

Jie Chen,[1,2] Kazuki Nitta,[2,3] Xin Zhao,[1] Takahiko Mizuno,[3,4,5] Takeo Minamikawa,[3,4,5] Francis Hindle,[6] Zheng Zheng,[1,7,*] And Takeshi Yasui[3,4,5,*]

[1]School of Electronic and Information Engineering, Beihang University, Beijing, 100083, China

[2]Graduate School of Advanced Technology and Science, Tokushima University, Tokushima 770-8506, Japan

[3]JST, ERATO MINOSHIMA Intelligent Optical Synthesizer (IOS), Tokushima 770-8506, Japan

[4]Institute of Post-LED Photonics, Tokushima University, Tokushima 770-8506, Japan

[5]Graduate School of Technology, Industrial and Social Sciences, Tokushima University, Tokushima 770-8506, Japan

[6]Laboratoire de Physico-Chimie de l'Atmosphère, Université du Littoral Côte d'Opale, Dunkerque 59140, France

[7]Beijing Advanced Innovation Center for Big Date-based Precision Medicine, Beihang University, Beijing, 100083, China

Abstract

Mode-resolved dual-comb spectroscopy (DCS) is an emerging spectroscopic tool with the potential to simultaneously achieve a broad spectral coverage and ultrahigh spectral resolution in terahertz (THz) spectroscopy. However, the need for two independently stabilized ultrafast lasers significantly hampers the potential application of DCS techniques. In this article, we demonstrate mode-resolved DCS in the THz region based on a free-running single-cavity dual-comb fiber laser with adaptive sampling. Low-pressure spectroscopy of acetonitrile gas with absorption features approaching the Doppler limit is demonstrated by comb-mode-resolved measurements with a spectral sampling spacing of 48.8 MHz, a spectral resolution of less than 5 MHz and a signal-to-noise ratio of ~50 dB. The successful demonstration of the proposed method clearly indicates the great potential for the realization of low-complexity, MHz-resolution THz spectroscopy instrumentation.

# 1. INTRODUCTION

Coherent spectroscopic techniques in the terahertz (THz) or far-infrared region (frequencies of 0.1 ~ 10 THz and wavelengths of 30 μm ~ 3 mm) are the enabling technology for a wide variety of important applications, ranging from material characterization for intriguing and highly complex interactions [1,2] to submillimeter-wave electronic device/system characterization [3]. Among these techniques, photonic-based THz time-domain spectroscopy (THz-TDS) has developed into arguably the most widely used technology in the past three decades [4,5]. THz-TDS takes advantage of the broadband radiation of ultrafast lasers and optically pumped THz emitters/receivers relative to its narrowband electronic-based counterparts to realize measurements over a large THz spectral range. THz-TDS has been applied to the study of a diverse range of samples, such as the rotational transitions of polar gas molecules [6], the hydrogen bonding signature in aqueous systems [7], and the dynamics and self-assembly of proteins [8]. This well-established THz-TDS scheme uses ultrashort pulses from a mode-locked laser for THz radiation and delayed probe pulses for THz detection; however, it has a limited spectral resolution and accuracy due to the constraints on the travel range, repeatability, and speed of the mechanical delay lines. To overcome this limitation, asynchronous optical sampling (ASOPS) THz-TDS [9-11] has been developed, using two mode-locked femtosecond lasers with a small repetition rate offset between them. By avoiding the physical delay lines, the equivalent temporal delay range can be largely increased; thus, the spectral resolution can be significantly enhanced to ~1 GHz. The emergence of optical frequency combs, which provide unprecedented freedom and accuracy in manipulating optical frequencies, also benefits the development of their counterparts in the THz region. THz dual-comb spectroscopy (THz-DCS) [12-14] has become a promising pathway towards ultrahigh-resolution broadband THz spectroscopy. While directly generating two THz combs is possible using a pair of THz quantum cascade lasers (QCLs) [15-17] for example, such THz sources have a relatively large comb tooth spacing and often poor mutual coherence between the combs in place of compact device size. This poses limitations to their spectral resolution and sampling spacing. If two stabilized optical combs with a high mutual stability are used for the generation and detection of a THz comb, the mixing of dual THz combs with excellent mutual coherence can be mapped to an RF comb with the temporal magnification factor (TMF), which is given by the ratio of the repetition rate to the repetition rate offset ($f_{rep1}/\Delta f_{rep}$). An RF comb is easily accessible with low-bandwidth electronics and further processed to yield high-bandwidth, high-resolution THz spectroscopic information. It has been demonstrated that the spectral resolution can reach the THz comb tooth linewidth, which is on the order of MHz or better [18,19]. By sweeping the lasers' comb tooth spacing via slight adjustments in the lasers' cavity lengths, gapless THz spectral sampling was achieved [20,21]. However, the practical use of mode-resolved THz-DCS is still hampered by the need for two frequency-stabilized optical frequency comb sources, which are expensive and complex.

Various schemes have been investigated to further reduce the complexity of THz-DCS systems and optical DCS systems. Recently, advances in the endeavor to generate a pair of frequency combs from a free-running single-cavity laser [22-31] have shown great potential towards this goal. By propagating through the same cavity, the dual optical combs experience almost the same disturbances, and the common-mode fluctuations thereby prevent the decline of the mutual coherence between the dual combs. Such a single-cavity dual-comb laser has been applied to THz and microwave frequency measurements [32,33] and optical spectroscopy [28,34-36]. Furthermore, a single-cavity dual-comb laser has been successfully demonstrated to realize low-complexity THz-DCS [37,38]; however, the spectral resolution remained at approximately 1 GHz and was insufficient for low-pressure, Doppler-limited gas spectroscopy. In this case, the long-term instability in the TMF caused by the residual timing jitter of the laser eventually limits the spectral resolution [39,40]. On the other hand, adaptive sampling has been shown to be able to compensate for the timing jitter in DCS in the optical and THz regions [41,42]. Adaptive sampling can reconstruct the relative coherence for relatively long time scales between the two combs. Furthermore, from the viewpoint of the suppressing the timing jitter, it has been demonstrated that a combination of adaptive sampling with two free-running lasers is more powerful than a combination of constant sampling with two stabilized lasers [42]. However, two similarly constructed ultrafast lasers are still required. A combination of adaptive sampling with a free-running single-cavity dual-comb laser will be the ultimate form of THz-DCS for high spectroscopic performance and reduced system complexity.

In this paper, we demonstrate a comb-mode-resolved THz-DCS scheme based on a simple free-running, single-cavity dual-comb fiber laser and adaptive sampling. By extending the temporal sampling window to over 200 ns while maintaining the data fidelity over a long acquisition period, a mode-resolved THz comb spectrum with a frequency sampling spacing of 48.8 MHz and a spectral resolution of less than 5 MHz is obtained with a signal-to-noise ratio (SNR) of ~ 50 dB.

## 2. METHODS

THz-DCS can be performed in the frequency domain [12,13] or time domain [14]. In time-domain THz-DCS, the THz frequency comb spectrum is obtained by a combination of ASOPS and a Fourier transformation (FT), as shown in Fig. 1(a). Using an ASOPS-THz-TDS system consisting of two mode-locked lasers with slightly mismatched repetition rates ($f_{rep1}$ and $f_{rep2}$, $\Delta f_{rep} = f_{rep2} - f_{rep1}$) for THz generation and detection, a THz pulse train (with a repetition rate of $f_{rep1}$, $N$ consecutive THz pulses and a time window size of $N/f_{rep1}$) is slowed down to an RF pulse train (with a repetition rate of $\Delta f_{rep}$, N consecutive RF pulses and a time window size of $N/\Delta f_{rep}$) by the TMF ($f_{rep1}/\Delta f_{rep}$) based on the temporally magnifying function of ASOPS; this enables a direct acquisition of the RF pulse train using a digitizer or an oscilloscope without the need for mechanical time-delay scanning. The FT of the RF pulse train results in the mode-resolved RF comb spectrum with a frequency spacing of $\Delta f_{rep}$. Finally, the mode-resolved THz comb spectrum is obtained by calibrating the frequency scale of the RF comb spectrum with the inverse of the TMF. The mode-resolved THz comb spectrum contains a series of successive narrow lines with a frequency spacing equal to the repetition rate ($f_{rep1}$) and a spectral sampling step equal to the inverse of the time window ($f_{rep1}/N$).

However, the RF pulse train is subject to the residual timing jitter of the laser source due to variations in the TMF, leading to a nonlinearity in the RF time scale. The fluctuating RF pulse train is shown in the upper row of Fig. 1(b) [42]. Such temporal fluctuations will be transferred to the frequency scale of the RF comb and THz comb, which will seriously degrade the spectral resolution and accuracy. If the RF pulse train is acquired by a sampling clock synchronized with such fluctuations as shown in the middle row of Fig. 1(b), the time-scale linearity of the sampled signal can be recovered (see the lower row of Fig. 1(b)). In this case, the time window size can be effectively extended without an accumulation of timing errors, and it is also possible to acquire accumulated data over a longer period of time to realize an improved SNR.

As shown in Fig. 2, the comb-mode-resolved adaptive sampling THz-DCS setup is seeded by a free-running single-cavity dual-comb fiber laser, which has an all-fiber ring cavity in which the dual-comb light beams propagate along a common-path route. The dual-comb light beams with different center wavelengths were obtained by multiplexing the mode-locking operation in the wavelength region [32,34,37]. The cavity consists of a hybrid wavelength division multiplexer and isolator (WDM/ISO), a 0.46-m-long erbium-doped fiber (EDF, Er110) pumped by a 980 nm laser diode, a polarization controller (PC), a single-wall carbon nanotube saturable absorber (SA), an in-line polarizer (ILP) with polarization-maintained fiber (PMF) pigtails and a 40 % fiber output coupler (OC). In addition, a 65-cm-long dispersion compensation fiber (DCF) is installed to optimize the intracavity dispersion at ~4.0 fs/nm at 1560 nm. The total length of the cavity is ~4.21 m. Due to the considerable birefringence of the PMF and the use of the ILP, spectral filtering and polarization-dependent loss tuning can enable dual-comb lasing generation by adjusting the PC [34].

The two-color dual optical combs from the laser are separated by a bandpass filter into two independent combs ($\lambda_1$-comb and $\lambda_2$-comb) and further amplified by erbium-doped fiber amplifiers (EDFAs) to pump other parts of the system. The THz comb is radiated from a fiber-coupled, strip-line-shaped LT-InGaAs/InAlAs photoconductive antenna PCA1 (TERA 15-TX-FC, Menlo Systems, bias voltage = 20 V, and optical power of 20 mW) excited by $\lambda_1$-comb pump light and then passes through a low-pressure gas cell (length = 38 cm and diameter =17 mm). The THz comb is then detected by another fiber-coupled, dipole-shaped LT-InGaAs/InAlAs photoconductive antenna PCA2 (TERA 15-RX-FC, Menlo Systems, optical power = 20 mW) that is pumped by the $\lambda_2$-comb light. The electrical output of PCA2 is amplified by a current preamplifier (AMP, bandwidth =3.8 MHz, and gain = $1\times10^6$ V/A). The temporal waveform of the output is acquired with a data acquisition board.

A portion of the separated $\lambda_1$-comb and $\lambda_2$-comb light is fed into a sum-frequency-generation cross-correlator (SFG-X), whose setup is constructed based on a noncollinear configuration with a piece of a β-BaB2O4 (BBO) crystal. The resulting SFG pulse that occurs every $1/\Delta f_{rep}$ serves as the trigger signal for the data acquisition board.

The other important part of the setup compared to our previous THz-DCS system [37] is the adaptive clock generator, which suppresses the long-term drift and timing jitter in the repetition rate difference Δfrep of the free-running, dual-comb laser. As shown in Fig. 2, the adaptive clock, which can trace the timing fluctuation in real time, is generated by heterodyning photoconductive mixing of reference CW-THz radiation and two PC-THz combs (PC-THz comb1 and PC-THz comb2) seeded by the optical dual combs [42]. In this scheme, two bowtie-shaped, low-temperature-grown GaAs photoconductive antennas (BT-PCA1 and BT-PCA2) are optically excited by the second-harmonic-generation (SHG) light of the optical dual combs using periodically poled lithium niobate (PPLN) crystals (not shown in Fig. 2). When the CW-THz radiation (fTHz = 0.1 THz, linewidth < 0.6 Hz, and average power = 2.5 mW) from an active frequency multiplier chain sourced by a microwave frequency synthesizer is also incident on both BT-PCA1 and BT-PCA2, two RF beat signals are generated between the CW-THz and the nearest adjacent PC-THz comb lines. Their frequencies ($f_{beat1}$ and $f_{beat2}$) are given by $|f_{THz} - mf_{rep1}|$ and $|f_{THz} - mf_{rep2}|$, respectively, assuming that the same m-order comb lines are involved. The $f_{beat2}$ - $f_{beat1}$ signal that

carries the timing fluctuation information (= $m\Delta f_{rep}$) is obtained by electrical mixing with a double-balanced mixer (M). The signal is further frequency-multiplied by a frequency multiplier (FM, frequency multiplication factor N = 40), which serves as the adaptive sampling clock of the acquisition board. The mode-resolved THz comb spectrum can be obtained by taking the Fourier transform of the temporal waveform accumulated in the time domain. A rubidium frequency standard (Stanford Research FS725, accuracy = $5\times10^{-11}$ and instability = $2\times10^{-11}$ at 1 s) is used to provide the common time-base signal for the CW-THz source and the data acquisition board (not shown in Fig. 2).

## 3. RESULTS

A. Performance of the dual-comb fiber laser

When the EDF is pumped above its mode-locking threshold and the intracavity PC is properly adjusted, dual-comb lasing is achieved with similar peak spectral intensities at two different center wavelengths, as shown in Fig. 3(a). The spectra at 1532.5 nm and 1557.7 nm have 3 dB bandwidths of 4.2 nm and 3.6 nm, respectively (namely, the $\lambda_1$-comb light and $\lambda_2$-comb light). The interval between their wavelengths is 25.2 nm, consistent with the birefringence introduced by the PMF fiber. Due to the relatively low intracavity anomalous dispersion achieved by optimizing the length of the DCF, the $\lambda_1$-comb light and $\lambda_2$-comb light have slightly different repetition rates, $f_{rep1}$ (~ 48.804486 MHz) and $f_{rep2}$ (~ 48.804296 MHz), and their difference ($\Delta f_{rep}$) is only ~190 Hz, as shown in Fig. 3(b). After separate amplification by EDFAs, the dual-comb lights are spectrally broadened and temporally compressed to ~110 fs at full-width-at-half-maximum (FWHM) by propagating through standard single-mode fiber (SMF), which are sufficient to drive the broadband THz comb spectrum.

The temporal drifts of $f_{rep1}$, $f_{rep2}$ and $\Delta f_{rep}$ under free-running conditions are monitored in time, since they affect the TMF between the THz comb and RF comb (see Fig. 1). As shown in Fig. 3(c), the drifts of both repetition rates are ~2.5 Hz in 120 s and follow the same trend owing to the identical fluctuations in the completely shared path for both pulses. The resulting fluctuation of $\Delta f_{rep}$ is merely 5.1 mHz in terms of standard deviation without active stabilization. The frequency instability of $f_{rep1}$ and $\Delta f_{rep}$ in the single-cavity dual-comb laser was measured for different gate times as indicated by the red solid and hollow circles in Fig. 3(d). For comparison, we also evaluated the frequency instability of $f_{rep1}$ and $\Delta f_{rep}$ for two independently stabilized fiber-comb lasers [see the blue solid and hollow circles in Fig. 3(d)] and two independently free-running fiber-comb lasers [see the green solid and hollow circles in Fig. 3(d)]. When comparing the single-cavity dual-comb laser with two independently free-running fiber-comb lasers, $f_{rep1}$ and $\Delta f_{rep}$ show similar trends. However, the frequency stability of $\Delta f_{rep}$ in the single-cavity dual-comb laser is significantly better than that in the two free-running fiber-comb lasers over a short time. Next, when comparing the single-cavity dual-comb laser with two independently stabilized fiber-comb lasers, the single-cavity dual-comb laser shows better short-term stability in $\Delta f_{rep}$ than the two independently stabilized fiber-comb lasers. On the other hand, the two independently stabilized fiber-comb lasers show better long-term stability in $\Delta f_{rep}$ and $f_{rep1}$ than the single-cavity dual-comb laser. Importantly, even though the single-cavity dual-comb laser is inferior to the two independently stabilized fiber-comb lasers in terms of the long-term stability of $\Delta f_{rep}$ and $f_{rep1}$ due to the lack of active laser stabilization, the adaptive sampling method allows us to overcome such inferiority significantly as demonstrated later, leading to a reduced system complexity and excellent spectroscopic performance.

B. Performance of the adaptive sampling scheme

To investigate the effectiveness of the proposed adaptive sampling THz-DCS method, 100,000 temporal waveforms of a train of 10 consecutive THz pulses were acquired and accumulated with the adaptive sampling clock method. For a comparison, similar temporal waveforms were acquired based on the conventional method with a constant sampling clock. As shown in the upper part of Fig. 4(a), when using the constant sampling method, the THz pulses almost disappeared except for the first pulse, because the residual timing jitter causes random walk-off of the temporal sampling positions in each time-delay scan and hence quite a low efficiency in the signal accumulation. Obviously, the constant sampling method is not suitable to extend the temporal window of the accumulated temporal waveform to multiple pulse periods in the case of a free-running single-cavity dual-comb laser. However, by using the adaptive sampling method, each THz pulse can be clearly observed in the accumulated temporal waveform, as shown in the lower part of Fig. 4(a). Moreover, the peak amplitude of the pulsed THz electric field remains constant for each THz pulse after many averages. These results imply that the adaptive sampling method has the capability to suppress the instability of the TMF over a long data acquisition time. Next, the comb-mode-resolved THz comb spectra were obtained by calculating the FT of the adaptive sampling temporal waveform [see the lower part of Fig. 4(a)], as shown in Fig. 4(b). Two spectral dips caused by atmospheric water vapor absorption appear at 0.557 THz and 0.752 THz with a pressure broadening linewidth of approximately 7 GHz (FWHM). An expanded view of the spectral region around 0.567 THz is shown in the inset of Fig. 4(b), indicating discrete lines with a constant frequency spacing of 48.8 MHz, which is consistent with the laser repetition rate. The spectral width of each comb line is less than ~5 MHz, which is exactly equal to the FT sampling resolution (1/10 of the comb mode spacing in our case). This spectrum clearly indicates that the spectral resolution of our setup can reach the MHz level or below without the influence of the residual timing jitter in the free-running single-cavity dual-comb laser. In this way, comb-mode-resolved spectroscopy is successfully demonstrated by the adaptive sampling THz-DCS system using just one simple fiber laser.

C. THz spectroscopy of a low-pressure molecular gas

To demonstrate the high-resolution spectroscopic capability of the proposed system, gas-phase THz spectroscopy of acetonitrile ($CH_3CN$) is performed at low pressure. Since $CH_3CN$ is an abundant species in the interstellar medium, a volatile organic gas, and even present in human breath and related to esophageal cancer, there is a considerable need for spectroscopic measurements of this molecule. $CH_3CN$ has relatively complicated spectral absorption features in the THz domain. $CH_3CN$ is a symmetric top molecule with a rotational constant $B$ of 9.194 GHz and a centrifugal distortion constant $D_{JK}$ of 17.74 MHz [43], and the frequencies of rotational transitions are given by

$$v = 2B(J+1) - 2D_{JK}K^2(J+1) \tag{1}$$

where $J$ and $K$ are rotational quantum numbers. There are two sets of characteristic features in the THz spectrum of $CH_3CN$. From the first term in Eq. (1), there are multiple manifolds of rotational transitions that are equally spaced by 2B (≈18.388 GHz). From the second term in Eq. (1), each manifold consists of a series of closely spaced absorption lines on the order of tens of MHz. The coexistence of GHz- and MHz-order absorption features makes this gas a good candidate for the demonstration of high-resolution spectroscopic measurements.

Mode-resolved THz comb spectra were obtained from the FT of temporal waveforms of a train of 10 consecutive THz pulses (accumulation number of 100,000). Figure 5(a) compares the logarithmic power spectra when the THz comb passed through a vacuum gas cell (blue plot) and a gas cell filled with a mixture of CH3CN and air with a total pressure of 360 Pa (red plot). The measured THz spectra display a power dynamic range of ~50 dB in the 0.1-1.5 THz frequency band. Two strong absorption lines appear at 0.557 and 0.752 THz in both spectral envelopes, originating from the atmospheric water vapor. Figure 5(b) shows a zoomed-in plot of the linear power spectra for Fig. 5(a) within the spectral range of 0.3305 to 0.3315 THz for the vacuum gas cell (blue plot) and the CH3CN/air gas cell (red plot). Although a periodic modulation is observed in the spectral envelope of the THz comb mode due to internal reflections in the THz optical setup, it can be canceled by calculating the absorption spectrum from both spectra. Because this frequency range has no absorption lines from water vapor, the difference in the spectra reflects the absorption features of the low-pressure $CH_3CN$ gas.

The spectrum of the absorption coefficient for $CH_3CN$ gas was obtained by normalizing the mode-resolved THz comb spectrum measured in the 360 Pa $CH_3CN$/air gas cell with that measured in the vacuum gas cell. Figure 6(a) shows the broadband spectrum of the absorbance coefficient, indicating multiple manifolds with a peak absorption coefficient of approximately 0.1 cm$^{-1}$. The manifolds were separated by ~18.388 GHz, exactly equal to *2B*, and a series of manifolds within this frequency range were correctly assigned to *J* = 10 to 38. Figure 6(b) shows a zoomed-in plot of Fig. 6(a) within the frequency range of 0.31 to 0.37 THz. In addition to J = 16 to 19 for the ground state, the manifolds of the rotational transitions for vibrationally excited states [21] were clearly observed as marked by the red asterisks. These spectral features were obtained by the enhanced spectral resolving power and sensitivity associated with the low-pressure gas condition, which constitutes a significant enhancement in the spectroscopic performance compared to the previous demonstration [37].

The observed manifolds consist of a number of closely spaced rotational transitions, assigned by K, as tabulated in the Jet Propulsion Laboratory (JPL) spectral database [44]. Multipeak fitting analysis was performed, in which a Lorentzian lineshape was used for each rotational line and a global linewidth parameter was applied to all the lines. The line positions and intensities were fixed parameters, while the acetonitrile partial pressure and linewidth were left as free parameters. An example is presented in the red and blue plots of Fig. 6(c), yielding a $CH_3CN$ partial pressure of 127 ($\pm$ 1) Pa and an FWHM linewidth of 77 ($\pm$ 2) MHz. The database fitting curve is in good agreement with the experimental data for all peaks in the frequency range of 0.2 to 0.4 THz with a slight residual [see the black plot of Fig. 6(c)]. The reason for the periodic residual is that the fitting model does not include rotational transitions for the vibrationally excited states [44].

Figures 6(d) and 6(e) show zoomed-in spectra for J = 17 at 0.3310 THz and J = 19 at 0.3677 THz, in which the blue plots show the experimental data. The frequency spacing of the experimental data points is ~48.804 MHz, which is equal to the comb mode spacing. The position and relative integrated intensities of the rotational absorption lines in the JPL spectral database are shown as the green dashed lines and the green plots together with the K number. The database fitting curve, shown by the red line in Figs. 6(d) and 6(e), is in good agreement with the experimental data. In this way, by using a free-running single-cavity dual-comb fiber laser combined with adaptive sampling, we successfully demonstrate high-resolution THz spectroscopy of low-pressure molecular gases.

An additional validation of our THz-DCS scheme was carried out by measuring the pressure broadening of the CH3CN absorption spectrum at different gas pressures. The comb-mode-resolved absorption spectra were measured around 0.331 THz by varying the pressure of the CH3CN/air mixture gas from 430 Pa, 330 Pa, 280 Pa, 256 Pa, 149 Pa to 115 Pa, as shown by the blue plots in Figs. 7(a) - (f). A comparison of these experimental data clearly indicates the pressure-dependent change in the shape of the absorption spectrum. To perform a quantitative analysis, we again performed multipeak fitting analysis based on a Lorentzian lineshape under the assumption that all K lines have the same linewidth. As shown by the red line in Fig. 7(a) - (f), the database fitting spectra match the experimental data very well. These analyses suggest that our THz-DCS scheme with a reduced system complexity has the capability to realize low-pressure gas spectroscopy with a MHz-order spectral resolution.

## 4. DISSUSSION

We first discuss the potential of the mode-resolved adaptive sampling THz-DCS scheme for low-pressure, Doppler-limited gas spectroscopy. We confirmed clear differences in the absorption linewidth as the pressure was reduced. By using multipeak curve fitting analysis based on the JPL spectral database, the results indicated that the observed pressure broadening coefficient lies between the values for self-broadening (912 kHz/Pa) and broadening by nitrogen (91.2 kHz/Pa) [45,46]. The Doppler-limited linewidth of $CH_3CN$ gas was 382 kHz (FWHM) at 0.2 THz and 1.34 MHz (FWHM) at 0.7 THz [47]. Consistent with these values, mode-resolved adaptive sampling THz-DCS can interrogate low-pressure gases with absorption features approaching the Doppler limit.

We next discuss the possibility of employing the proposed method for THz spectroscopy with improved precision. Although we demonstrated the potential of the proposed method for Doppler-limited, low-pressure gas spectroscopy, the frequency comb spacing of $f_{rep1}$ is still coarse for a full analysis of each rotational transition with a MHz-order structure. A promising approach for THz spectroscopy with improved precision is to use the gapless technique in the THz comb, in which the frequency spacing of two stabilized THz combs is precisely swept to interleave additional comb lines into the original comb lines [20, 21]. In this case, the THz comb spacing can be reduced down to the linewidth of the THz comb mode from $f_{rep1}$. To expand this technique into the free-running THz combs, we have to consider the frequency instability of the THz comb line because it determines the spectrally interleaved interval. Although the frequency instability of the THz comb line depends on both the fluctuations of $\Delta f_{rep}$ and $f_{rep1}$, the former can be well compensated by the adaptive sampling method, as shown in Fig. 4(a). The fluctuation of frep1 is a remaining factor associated with the fluctuation of the THz comb line position. As the frequency instability of $f_{rep1}$ was approximately $10^{-8}$ from Fig. 3(d), it is expected that the THz comb line fluctuated within a frequency range of 3.3 kHz at 0.33 THz and 5.5 kHz at 0.55 THz. Since this kHz fluctuation is much smaller than the MHz comb linewidth determined by the inverse of the time window size (4.8 MHz in this article), the influence of the fluctuating $f_{rep1}$ is negligible for gapless THz-DCS. frep1 and frep2 will be tuned with little change in $\Delta f_{rep}$ by mechanically stretching a portion of the cavity fiber with an additional piezoelectric actuator or motor-driven translation stage. The combination of mode-resolved adaptive sampling THz-DCS with the gapless technique will be studied in a future work.

## 5. CONCLUSIONS

We demonstrated the capability of realizing mode-resolved THz comb spectroscopy with a free-running single-cavity dual-comb fiber laser. By using adaptive sampling with the free-running laser, the long-term instability of the TMF was effectively suppressed, facilitating the long-term acquisition and temporal accumulation of THz temporal waveforms with a time window extending to multiple laser pulse periods. This results in a broadband, mode-resolved THz comb spectrum with a frequency sampling spacing of 48.804 MHz and a spectral resolution of less than 5 MHz. Low-pressure CH3CN/air gas with absorption spectral features on the order of GHz and MHz was measured, and good agreement with the theoretical predictions was achieved. The resolving capability of MHz-level spectral characteristics using a simple fiber laser could greatly expand the applicability of precise THz spectroscopic techniques to much broader areas.

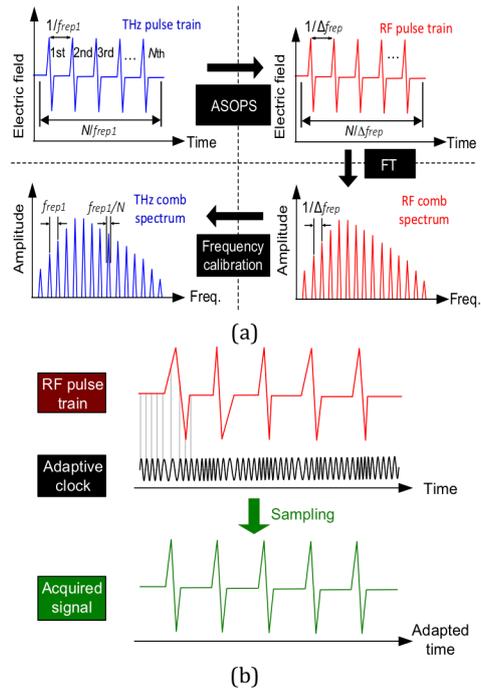

Fig. 1. Principle of operation. (a) Flowchart of time-domain THz-DCS. (b) Acquisition of the temporal waveform using the adaptive sampling method.

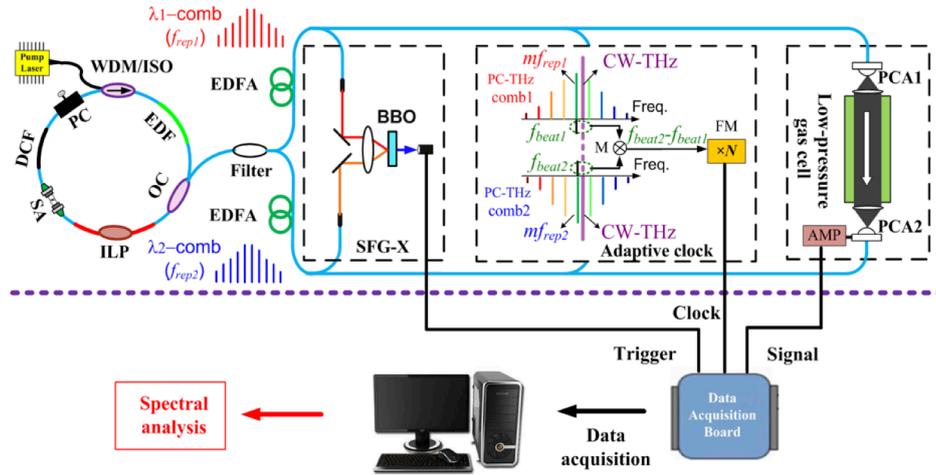

Fig. 2. Configuration of comb-mode-resolved adaptive sampling THz-DCS. SFC-X, sum-frequency-generation cross-correlator. BBO, beta-barium borate crystal. PC-THz comb, photocarrier THz comb. M, double-balanced mixer. FM, frequency multiplier (frequency multiplication factor=N= 40). PCA, photoconductive antenna. AMP, current preamplifier.

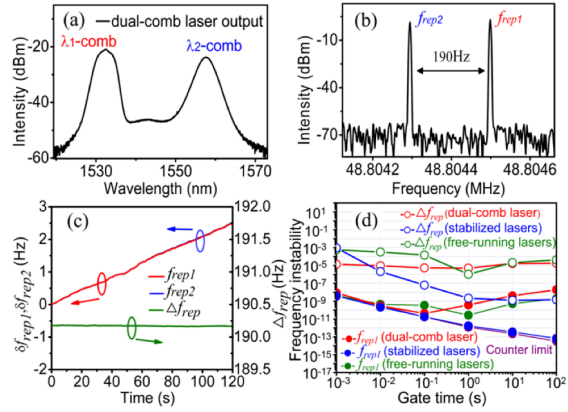

Fig. 3. Performance of the dual-comb fiber laser. (a) Output spectrum of the laser. (b) RF spectrum of the dual-comb pulses. (c) Fluctuations of $f_{rep1}$, $f_{rep2}$ and $\Delta f_{rep}$. (d) Measured frequency instability of $f_{rep1}$ and $\Delta f_{rep}$.

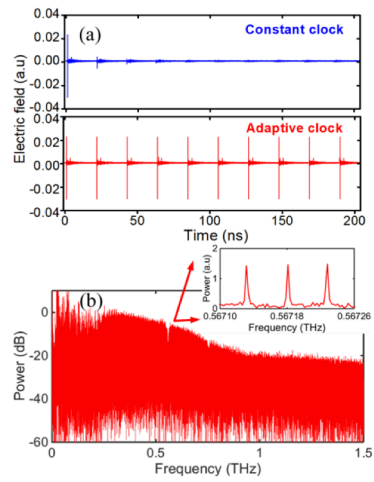

Fig. 4. (a) Comparison of the ASOPS waveforms averaged 100,000 times obtained using different sampling clocks. (b) Comb-mode-resolved THz spectrum through air at room pressure. Inset: a zoomed-in plot around 0.5672 THz.

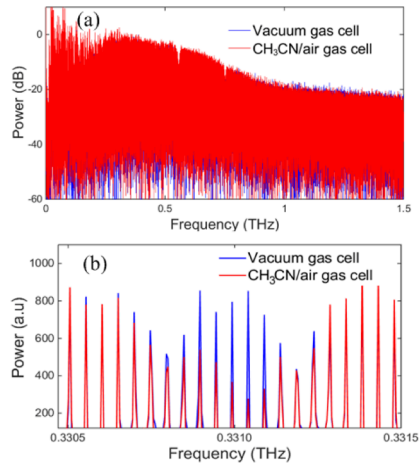

Fig. 5. Mode-resolved THz comb spectra of the vacuum gas cell (blue plot) and the 360 Pa $CH_3CN$/air gas cell (red plot). (a) Logarithmic power spectra within the frequency range of 0.1 to 1.5 THz and (b) zoomed-in plot of the corresponding linear power spectra (frequency range of 0.3305 ~ 0.3315 THz).

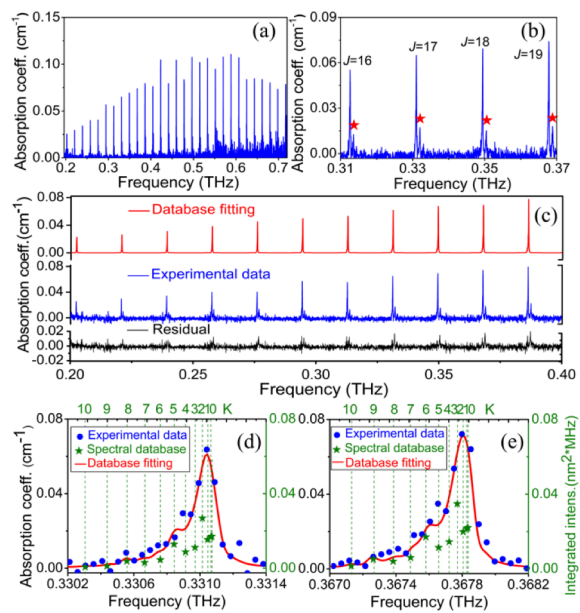

Fig. 6. Comb-mode-resolved THz spectroscopy of a mixture gas sample of $CH_3CN$ and air with a total pressure of 360 Pa. Absorption spectra of $CH_3CN$ within the frequency range of (a) 0.2 to 0.72 THz and (b) 0.31 to 0.37 THz. The red asterisks indicate the manifolds of the rotational transitions for the vibrationally excited states. (c) Comparison of the absorption spectra between the database fitting and the experimental data and their residual. (d) Absorption spectra around 0.3310 THz and (e) 0.3677 THz.

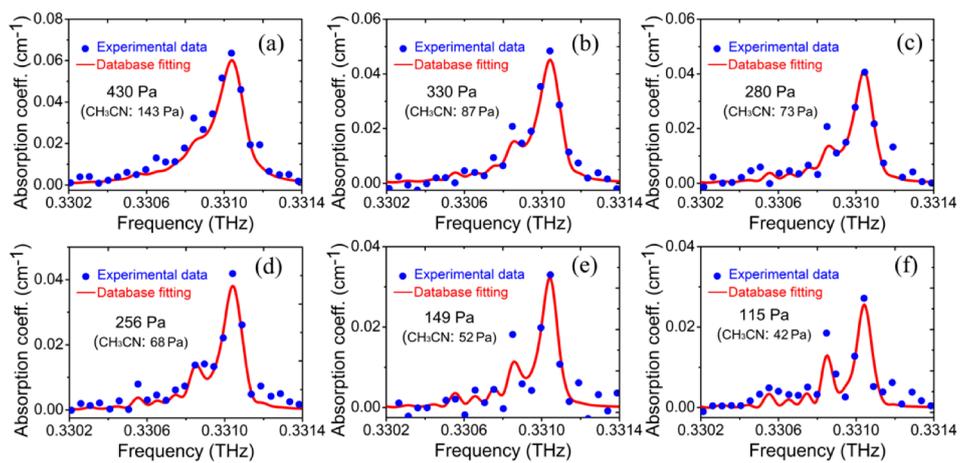

Fig. 7. Mode-resolved absorption characterization of $CH_3CN$ around 0.331 THz at (a) 430 Pa, (b) 330 Pa, (c) 280 Pa, (d) 256 Pa, (e) 149 Pa, and (f) 115 Pa.